\documentclass[a4paper,USenglish,cleveref]{lipics-v2021}

\usepackage{booktabs}
\usepackage{listings}
\usepackage{color}
\usepackage{bm}
\usepackage{multirow}
\usepackage{paralist}
\usepackage{xspace}
\usepackage{subcaption}

\renewcommand\vec{\mathbf}

\newcommand{\gggn}{{\it R-GNN}\xspace}
\newcommand{\cgnn}{{\it C-GNN}\xspace}
\newcommand{\ggat}{{\it R-GAT}\xspace}
\newcommand{\gggnnef}{{\it R-GNN}$_{\mbox{\footnotesize {\it NEF}}}$\xspace}
\newcommand{\ggatnef}{{\it R-GAT}$_{\mbox{\footnotesize {\it NEF}}}$\xspace}
\newcommand{\gggnns}{{\it R-GNN}$_{\mbox{\footnotesize {\it NS}}}$\xspace}
\newcommand{\gggnctx}{{\it R-GNN}$_{\mbox{\footnotesize {\it CTX}}}$\xspace}
\newcommand{\gggnnsctx}{{\it R-GNN}$_{\mbox{\footnotesize {\it NS-CTX}}}$\xspace}

\newcommand{\gnn}{GNN\xspace}
\newcommand{\gru}{GRU\xspace}

\newcommand{\deeptyper}{DeepTyper\xspace}
\newcommand{\lambdanet}{LambdaNet\xspace}

\newcommand{\identnode}{\textit{IdentNode}\xspace}
\newcommand{\toknode}{\textit{TokNode}\xspace}
\newcommand{\expnode}{\textit{ExprNode}\xspace}
\newcommand{\varsymnode}{\textit{VarSymNode}\xspace}
\newcommand{\objpropnode}{\textit{ObjPropNode}\xspace}
\newcommand{\ctxnode}{\textit{CtxNode}\xspace}

\newcommand{\expedge}{\textit{ExpEdge}\xspace}
\newcommand{\varsymedge}{\textit{VarSymEdge}\xspace}
\newcommand{\objpropedge}{\textit{ObjPropEdge}\xspace}
\newcommand{\retedge}{\textit{RetEdge}\xspace}
\newcommand{\calledge}{\textit{CallEdge}\xspace}
\newcommand{\ctxedge}{\textit{CtxEdge}\xspace}

\newcommand{\REM}[1]{}

\newcommand{\REV}[1]{#1}

\title{Advanced Graph-Based Deep Learning for Probabilistic Type Inference}

\author{Fangke Ye}
{Georgia Institute of Technology, USA}
{yefangke@gatech.edu}
{}{}

\author{Jisheng Zhao}
{Georgia Institute of Technology, USA}
{jisheng.zhao@cc.gatech.edu}
{}{}

\author{Vivek Sarkar}
{Georgia Institute of Technology, USA}
{vsarkar@gatech.edu}
{}{}

\authorrunning{F. Ye, J. Zhao, and V. Sarkar}



\nolinenumbers 

\hideLIPIcs  

\EventEditors{John Q. Open and Joan R. Access}
\EventNoEds{2}
\EventLongTitle{42nd Conference on Very Important Topics (CVIT 2016)}
\EventShortTitle{CVIT 2016}
\EventAcronym{CVIT}
\EventYear{2016}
\EventDate{December 24--27, 2016}
\EventLocation{Little Whinging, United Kingdom}
\EventLogo{}
\SeriesVolume{42}
\ArticleNo{23}

\begin{document}

\maketitle

\begin{abstract}
Dynamically typed languages such as JavaScript and Python have emerged as the most popular programming languages in use today.  However, when possible to do so, there are also important benefits that accrue from including static type annotations in dynamically typed programs, e.g., improved documentation, improved static analysis of program errors, and improved code optimization.  This approach to gradual typing 
is exemplified by the TypeScript programming system which allows programmers to specify partially typed programs, and then uses static analysis to infer as many remaining types as possible.  However, in general, static type inference is unable to infer all types in a program; and, in practice, the effectiveness of static type inference depends on the complexity of the program's structure and the initial types specified by the programmer.  As a result, there is a strong motivation for new approaches that can advance the state of the art in statically predicting types in dynamically typed programs, and that do so with acceptable performance for use in interactive programming environments.

Previous work has demonstrated the promise of {\em probabilistic type analysis} techniques that use deep learning methods such as recurrent neural networks  and graph neural networks (GNNs) to predict types for variable declarations and occurrences.  
In this paper, we advance past work by introducing a range of 
graph-based deep learning models that operate on a novel \textit{type 
flow graph} (TFG) representation.  The TFG represents an input program's elements as graph nodes connected with syntax edges and over-approximated data flow edges, and our GNN models are trained to predict the type labels in the TFG for a given input program.

We study different design choices for our GNN-based type inference system for the 100 most common types in our evaluation corpus, and show that our best GNN configuration for accuracy (\gggnnsctx{}) achieves a top-1 accuracy of 87.76\%.  This outperforms the two most closely related deep learning type inference approaches from past work -- DeepTyper with a top-1 accuracy of 84.62\%  and LambdaNet with a top-1 accuracy of 79.45\%.  Alternatively, we can state the error (100\% - accuracy) for \gggnnsctx{} is 0.80$\times$ that of DeepTyper and 0.60$\times$ that of LambdaNet.  Further, the average inference throughput of \gggnnsctx{}  is 353.8 files/second, compared to  186.7 files/second for DeepTyper and  1,050.3 files/second for LambdaNet.  If inference throughput is a higher  priority, then the recommended model to use from our approach is the next best GNN configuration from the perspective of accuracy (\gggnns{}) which achieved a top-1 accuracy of 86.89\% and an average inference throughput of 1,303.9 files/second.
In summary, our work introduces advances in graph-based deep learning that yield superior accuracy and performance to past work on  probabilistic type analysis, while also providing a range of GNN models that could be applicable in the future to other graph structures used in program analysis beyond the TFG.

\end{abstract}

\section{Introduction}\label{sec:intro}
Dynamically typed languages such as JavaScript and Python have emerged as some of the most popular programming languages in use today, as evidenced by their popularity ranks on GitHub.%
\footnote{The State of the Octoverse: \url{https://octoverse.github.com/}.}
Compared with statically typed programs, dynamically typed programs are usually 
more succinct and easier to modify.
However, when possible to do so, there are also important benefits
that accrue from including static type annotations in dynamically
typed programs, e.g., improved documentation and static analysis
of program errors.
The classical approach to address this problem is to perform {\em static type inference}, usually 
accomplished by applying control and data flow analyses to infer the relevant type lattices for 
expressions in a program.  However, in general, static type inference
is unable to infer all types in a dynamically typed program; and, in practice, the
effectiveness of static type inference depends on the complexity of
the program's structure, e.g., use of  implicit polymorphic variables,
object properties, and dynamic 
type evaluation.
An alternate {\em gradual typing} approach
is exemplified by the TypeScript programming system which allows
programmers to specify partially typed programs, and then uses static
analysis to infer as many remaining types as possible.  While an
improvement over a fully static approach, gradual typing is also
encumbered by the inherent challenges faced by static analysis.
As a result, there is a strong motivation for new approaches that can advance the state of the art in statically predicting types in dynamically typed programs, and that do so with acceptable performance for use in interactive programming environments.

Previous work has demonstrated the promise of {\em probabilistic type
  analysis} techniques that use deep learning methods to predict
types for variable declarations and occurrences.
One approach from past work, DeepTyper, formulates type prediction as a sequence tagging 
problem and trains a recurrent neural network (RNN) model to solve
that problem~\cite{deeptyper}. Another approach from past work, LambdaNet, 
employs an auxiliary analysis to infer type dependences between program elements and extract them into a graphical representation, on which it trains a graph neural network (GNN) model to perform type prediction~\cite{lambdanet}.
The introduction of these approaches is a natural follow-on to work
from the past two decades on applying machine learning to address a
number of compilation-related problems, and that has been enabled by
the rapid growth of open source online code repositories such as the open-source projects hosted on
GitHub.
Examples of programming tasks that have been aided by the use
of machine learning include property prediction, code search, anomaly detection, and code completion~\cite{mlcodesurvey}.

In this paper, we advance past work by introducing a GNN-based type inference system consisting of a range of GNN models operating on a novel \textit{type 
flow graph} (TFG) representation. The TFG represents an input
program's elements as graph nodes connected with syntax edges and
over-approximated data flow edges, and our GNN models are trained to
predict type labels from the TFG of a given input program.
The TFG is efficient to construct, and can be built via simple bottom-up traversals of the program's abstract syntax tree. Our GNN models learn to propagate type information on the graph and predict types for the graph nodes.
We study different design choices for our GNN-based type inference system for the 100 most common types in our evaluation corpus, and show that our best GNN configuration for accuracy (\gggnnsctx{}) achieves a top-1 accuracy of 87.76\%.  This outperforms the two most closely related deep learning type inference approaches from past work -- DeepTyper with a top-1 accuracy of 84.62\%  and LambdaNet with a top-1 accuracy of 79.45\%.  Alternatively, we can state the error (100\% - accuracy) for \gggnnsctx{} is 0.80$\times$ that of DeepTyper and 0.60$\times$ that of LambdaNet.  Further, the average inference throughput of \gggnnsctx{}  is 353.8 files/second, compared to  186.7 files/second for DeepTyper and  1,050.3 files/second for LambdaNet.  If inference throughput is a higher  priority, then the recommended model to use from our approach is the next best GNN configuration from the perspective of accuracy (\gggnns{}) which achieved a top-1 accuracy of 86.89\% and an average inference throughput of 1,303.9 files/second.
Thus, our work introduces advances in graph-based deep learning that yield superior accuracy and performance to past work on probabilistic type analysis, while also providing a range of GNN models that could be applicable in the future to other graph structures used in program analysis beyond the TFG.

A summary of the contributions of this paper is as follows:
\vspace{-15pt}
\begin{itemize}
\item We propose a probabilistic analysis framework for type
inference based on {\em a family of GNNs}.
\item We introduce a lightweight {\em type flow graph} structure that
efficiently captures type flow information from a program's abstract syntax tree.
\item We propose several design choices for our family of GNNs, and
evaluate their accuracy and efficiency as described above.
\item We compare our approach with two state-of-the-art deep
learning based approaches for probabilistic type inference 
(DeepTyper and LambdaNet) and show that our approach yields superior
accuracy and performance to both past approaches.
\end{itemize}

The rest of this paper is organized as follows.
\cref{sec:back} provides background on type inference and graph neural networks. \cref{sec:method} introduces our GNN-based type inference framework, including graph construction and model architecture.
In \cref{sec:expr}, we evaluate our approach with different design choices and compare it with 
previous work. \cref{sec:relwork} presents related
work, and \cref{sec:conc} concludes the paper.

\section{Background}\label{sec:back}
\subsection{Static Type Inference}
\label{sec:sati}
The goal of type inference is to automatically deduce the type of an expression in a program. The deduced type can be either partially or fully, depends on the simplicity of the target programming language or the robustness of the type inference analysis.

Static type inference algorithms infer type information via solving the type constraints collected from the target program. A type constraint represents the aggregation and reduction of the type information between expressions in the program.
This allows type inference to be formulated as a graph analysis problem.

\subsection{Graph Neural Networks}
\label{sec:gnn}
Graph neural networks (GNNs) are machine neural network models that directly operate on graph-structured data.
A graph can be defined as $G = (V, E)$, where $V$ is the set of nodes and $E$ is the set of edges. The neighborhood of a node $v$ is defined as $\mathcal{N}(v) = \{u \in V ~|~ (u, v) \in E\}$.
A node $v$ can have attributes represented as a feature vector $\vec{x}_v$, and similarly, an edge 
$(u,v)$ can have attributes represented as a feature vector $\vec{e}_{uv}$. GNNs learn the 
representations of graph structures 
by propagating the states of nodes to their neighbors iteratively and transforming the nodes states 
into output representations.
Let $\vec{h}_v$ be the state vector of node $v$ 
and $K$ be the number of propagation steps,%
\footnote{In this paper, we use the term {\it propagation step} to represent the computation in an iteration. A propagation step is also called a {\gnn~ layer} in the literature.}
a general form of a \gnn can be defined as:%
\footnote{Although there exists a more general form that allows edge features to be updated during the propagation~\cite{rdg}, the one we present here is sufficient to formulate all the GNN architectures we will discuss in this paper.}
\begin{flalign}
&&\vec{h}_v^{(0)} &= \vec{x}_v && v \in V
\label{eq:gnn_init}\\
&&\vec{m}_{uv}^{(k)} &= f_{\mathrm{message}}^{(k)} \left( \vec{h}_u^{(k-1)},\vec{e}_{uv} \right) && k=1..K, v \in V, u \in \mathcal{N}(v)
\label{eq:gnn_msg}\\
&&\vec{a}_v^{(k)} &= f_{\mathrm{aggregate}}^{(k)} \left( \vec{h}_v^{(k-1)}, \left\{ m_{uv}^{(k)} ~\middle|~ u \in \mathcal{N}(v) \right\} \right) && k=1..K, v \in V
\label{eq:gnn_agg}\\
&&\vec{h}_v^{(k)} &= f_{\mathrm{update}}^{(k)} \left( \vec{h}_v^{(k-1)}, \vec{a}_v^{(k)} \right) && k=1..K, v \in V
\label{eq:gnn_update}
\end{flalign}

The state of each node  $\vec{h}_v$ is initialized with its feature vector $\vec{x}_v$ in 
\cref{eq:gnn_init}. At step $k$, a node receives messages from its neighbors through corresponding 
edges. A message vector $\vec{m}_{uv}$ is generated from a neighbor's hidden state $\vec{h}_u^{(k-1)}$ 
at step $(k-1)$ and the feature vector of the edge where the message is passed through (i.e., 
$\vec{e}_{uv}$), using the message function $f_{\mathrm{message}}^{(k)}$ in \cref{eq:gnn_msg}. The 
messages are then aggregated to a single vector using the aggregation function 
$f_{\mathrm{aggregate}}^{(k)}$ in \cref{eq:gnn_agg}. At the end of step $k$, the node's state 
$\vec{h}_v^{(k)}$ is updated from its state at the previous step $\vec{h}_v^{(k-1)}$ and the 
aggregated message vector $\vec{a}_v^{(k)}$ using the update function $f_{\mathrm{update}}^{(k)}$
in \cref{eq:gnn_update}.

The functions $f_{\mathrm{message}}^{(k)}$, $f_{\mathrm{aggregate}}^{(k)}$, $f_{\mathrm{update}}^{(k)}$ 
are neural networks containing trainable parameters. Those functions can be the same and share 
parameters across different steps (i.e., 
$f_{\_}^{(1)}=f_{\_}^{(2)}=\ldots=f_{\_}^{(K)}$) or they can be 
different functions and use a separate set of parameters for each step. Following the taxonomy 
in~\cite{gnnsurvey}, we use the term \textit{recurrent graph neural networks} to refer to GNNs with 
shared functions and parameters for all steps, and \textit{convolutional graph neural networks} to 
refer to GNNs with separate functions and parameters for different steps.

\subsection{Machine Learning Assisted Static Analysis}
\label{sec:saml}
There has been a large body of work that leverages machine learning to assist static analysis.
By their tasks, those approaches can be categorized into the following categories:

\textbf{\textit{Program Property Prediction.}}
In this class of tasks, the goal is to assign property labels to (part of) a program. The typical approach is to apply supervised learning to build a model that predicts the property labels from a program's representation.
The program representations that have been adopted by previous work include token sequences~\cite{deeptyper}, abstract syntax trees~\cite{code2vec,jsnice}, and some more complex graphs~\cite{Allamanis17,code2inv}.

\textbf{\textit{Learning to Parametric Program Analysis.}}
To make a trade-off between precision and cost, parametric analyses have been applied in real-world solutions~\cite{zhang13, oh2014}.
Unlike manual parameterization and heuristic parameter search, which rely on the expertise of users or heuristic developers, statistical learning approaches can learn from automatically generated data and have shown a good capacity for finding the optimal configuration for parameters including the granularity of program abstraction~\cite{liang11a, oh2015}.

\textbf{\textit{Program Specification Identification.}}
This scenario focuses on mining program specification from code corpora, including the 
API specifications~\cite{salento, taintapis}, the behaviors of a library~\cite{libpta} and code 
idioms~\cite{idiom}. These techniques can be applied for detecting API misuses, performing code 
completion, etc.

In this paper, we focus on type inference, which can be categorized as a program property prediction problem. We 
investigate the approach of using graph structure to present the input program and building 
machine learning models to predict program properties. A ~\gnn-based learning framework is 
designed to learn from graph-structured data extracted from code and predict types for program elements.

\section{Methodology}\label{sec:method}
This section describes our methodology that uses GNN based machine learning framework to perform probabilistic type inference.
Section~\ref{sec:framework} describes the problem statement and the proposed learning framework.
Section~\ref{sec:graph} gives the definition of the graph we use as input to our GNN models.
The design of the GNN architecture is introduced in Section~\ref{sec:learngnn}.

\subsection{Problem Statement}
\label{sec:framework}

\begin{figure}[h]
\centering
\begin{subfigure}[t]{\linewidth}
\includegraphics[width=\linewidth]{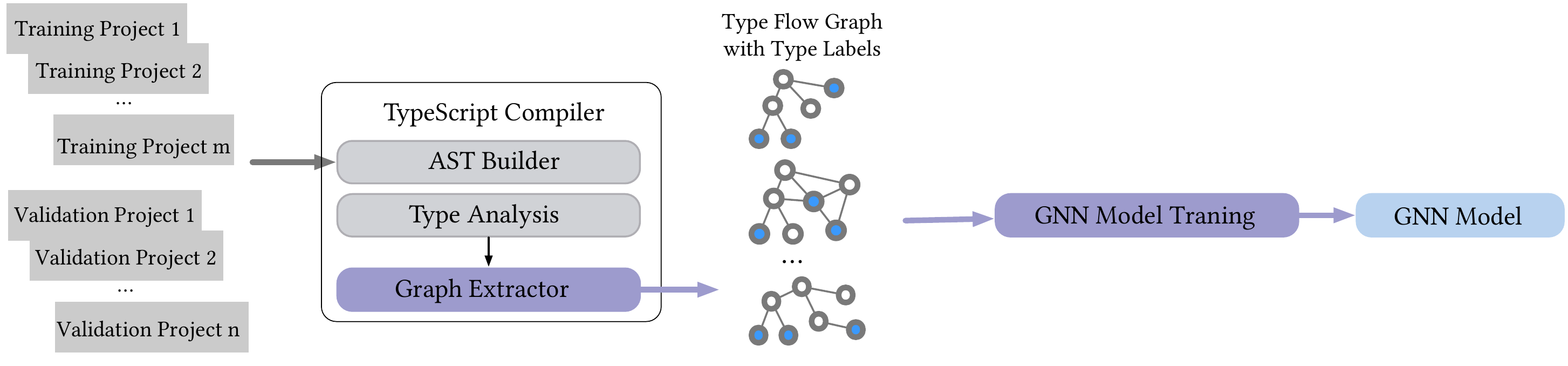}
\caption{Model training.}
\end{subfigure}
\par\medskip
\begin{subfigure}[t]{\linewidth}
\includegraphics[width=\linewidth]{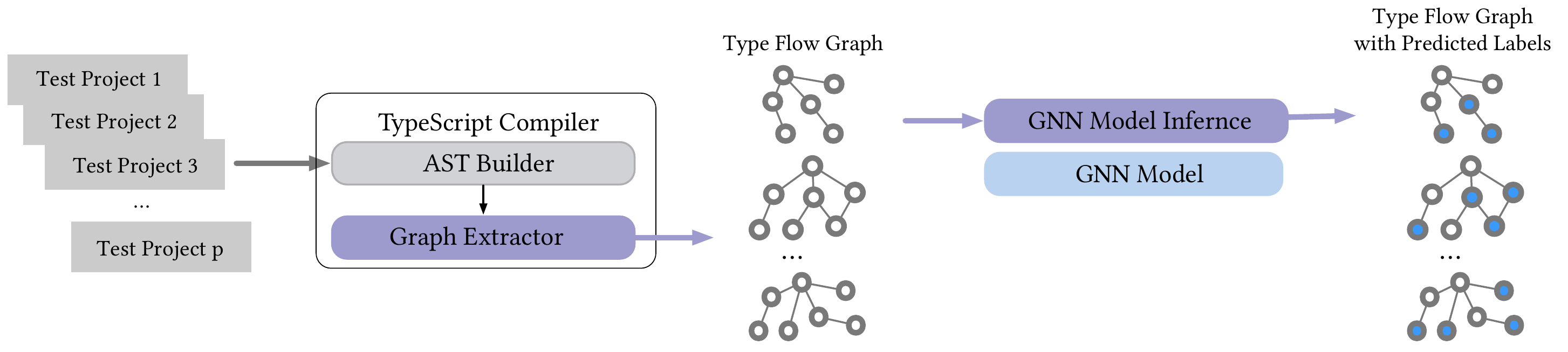}
\caption{Type inference.}
\end{subfigure}
\caption{The graph neural network based type inference framework.}
\label{fig:framework}
\end{figure}

We formulate type inference as a graph node label prediction problem.
The prediction is at the file level for better scalability and flexibility.
Our GNN-based type inference framework is shown in \cref{fig:framework}.
For each source file, we construct a graph whose nodes correspond to program elements in the file.
The graph edges include syntax edges extracted from the abstract syntax tree and over-approximated data flow edges derived from name-matching-based static analysis.
A GNN model is trained to predict type for the graph nodes
from a fixed type vocabulary.
The target programming language we choose is JavaScript, and the unit of program element for prediction is the identifier, which covers variables, function parameters, and object properties in JavaScript.
Following the methodology of previous work~\cite{deeptyper},
we train the model on a dataset of TypeScript programs.
The type labels are obtained from the TypeScript compiler, and the graphs are constructed from the code without using any information from existing type annotations.

\subsection{Graph Definition}
\label{sec:graph}
A data-driven approach relies on high-quality data containing features that are relevant to the target problem. For the type inference problem, the relevant features include the source of type information (e.g., literals), the code patterns that have implications about types, and the paths of type propagation.
Ideally, the input data to the learning model should at least cover those features. In this work, we defined a \textit{type flow graph} (TFG) that carries those features mentioned above by encoding syntactic and approximate semantic information in a program.

A TFG is a directed graph whose nodes represent program elements in the code and edges encode type-relevant relationships between the nodes. It is constructed from a program's abstract syntax tree (AST). We give the definitions of the nodes and edges and how they can be extracted in the following paragraphs.

\subsubsection{Graph Node}
The nodes in a TFG present the program elements that either carry types or provide hints about types.
They are categorized into the following node types:
\begin{itemize}
  \item {\it Identifier node} (\identnode):
  It represents an identifier token in a program and corresponds to an occurrence of a named element in the program, including variables, object properties, and functions.
  \item {\it Token node} (\toknode):
  It represents a non-identifier token that appears as leaves in an AST. Most of the token nodes in a program are literals.
  \item {\it Expression node} (\expnode):
  It represents an expression in a program, including binary, unary expressions, function and variable declarations;
  \item {\it Variable symbol node} (\varsymnode):
  It represents a variable and its occurrences within the live range; 
  \item {\it Object property node} (\objpropnode):
  It represents an object property name in a program. 
  It can be viewed as an over-approximation of an object property symbol;%
  \footnote{This is equivalent to applying 
  a flow-insensitive and context-insensitive alias analysis for object properties, which is over-approximated 
  since it only checks if properties' names are the same.}
  \item {\it Context node} (\ctxnode):
  It represents the context where an expression appears.
  Every \ctxnode is associated with an expression node in the AST whose parent is a statement node.
\end{itemize}

Each node has an associated feature. An \identnode's feature is its name and a \toknode's feature is its token type.
A \ctxnode's feature is a (\texttt{statement\_type}, \texttt{child\_name}) pair,
where \texttt{statement\_type} is the type of a statement node in the AST and \texttt{child\_name} 
is the name of a child node of that statement node (e.g., for an if-statement node with an 
expression child node serving as its condition, the expression node in the TFG will be connected to a \ctxnode with the 
feature (\texttt{IfStmt}, \texttt{condition})).
The features of other nodes are their node types.

\identnode{}s and \expnode{}s are the units for type prediction. \toknode{}s serve as sources of 
type information since they contain literals that have known types. A \varsymnode acts as a 
``hub'' that helps the type information exchange between different occurrences of the same variable.
And an \objpropnode has a similar functionality for object properties that share the same name.
A \ctxnode provides hints about the type of an expression based on its role in the enclosing 
statement.

\subsubsection{Graph Edge}
The functionality of graph edges is to establish information flow paths between the nodes.
To enable efficient information exchange, we make all edges bi-directional, and
each edge type we describe below has a corresponding backward edge type
in order to distinguish between a forward edge and a backward edge.
The edges are categorized into the following types:
\begin{itemize}
  \item {\it Expression edge} (\expedge):
  It connects an \identnode/\toknode/\expnode to an \expnode. The source and the destination of this edge should correspond to a child-parent pair in the AST.
  \item {\it Variable symbol edge} (\varsymedge):
  It connects an \identnode to a \varsymnode. The \identnode must represent a variable whose symbol corresponds to the \varsymnode.
  \item {\it Object property edge} (\objpropedge):
  It connects an \identnode to an \objpropnode. The \identnode must represent an object property whose name corresponds to the \objpropnode.
  \item {\it Return edge} (\retedge):
  It connects an \expnode representing a returned expression to an \expnode representing its enclosing function declaration.
  \item {\it Call edge} (\calledge):
  It can connect an \expnode representing a returned expression to an \expnode representing a call expression which is a potential caller of the enclosing function of that returned expression.
  It can also connect an \expnode representing an argument in a call expression to an \expnode representing a parameter of a function which is a potential callee of that call expression.
  \item {\it Context edge} (\ctxedge):
  It connects a \ctxnode to an \expnode it is associated with.
\end{itemize}

Each edge has an associated feature. The feature of an \expedge, presented as (\texttt{c}, \texttt{p}), and its backward edge, (\texttt{p}, \texttt{c}), is a (\texttt{expression\_type}, \texttt{child\_name}, \texttt{direction}) tuple,
where \texttt{expression\_type} is the expression type of \texttt{p}, \texttt{child\_name} is the child name of \texttt{c} in the AST, and \texttt{direction} is either \texttt{f} for a forward edge and \texttt{b} for a backward edge.
For other edges, their features are the edge types.

\begin{figure}[htbp]
\centering
\begin{subfigure}[t]{\linewidth}
\centering
\captionsetup{justification=centering}
\begin{tabular}{c}
\begin{lstlisting}[
language=C,
keywords={function,if,return,let,true},
basicstyle=\ttfamily\footnotesize,
keywordstyle=\color{blue},
stringstyle=\color{red},
backgroundcolor=\color{white}
]
function foo(a) {
  if (a.val) x = "Hello";
  return x;
}
r.val = true;
let c = foo(r);
\end{lstlisting}
\end{tabular}
\caption{Code example.}
\label{fig:tfg_code}
\end{subfigure}
\par\vspace{30pt}
\begin{subfigure}[t]{\linewidth}
\centering
\captionsetup{justification=centering}
\includegraphics[width=\linewidth]{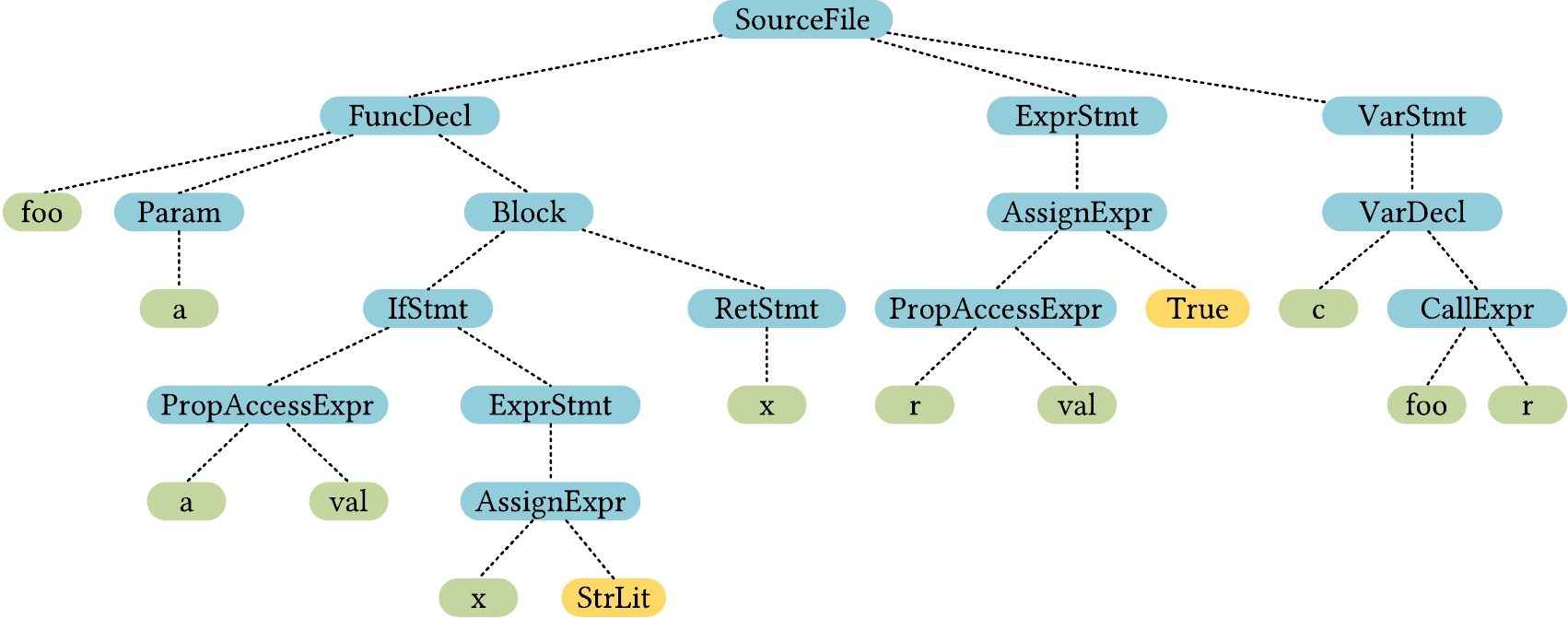}
\caption{Abstract syntax tree.}
\label{fig:tfg_ast}
\end{subfigure}
\par\vspace{30pt}
\begin{subfigure}[t]{\linewidth}
\centering
\captionsetup{justification=centering}
\includegraphics[width=\linewidth]{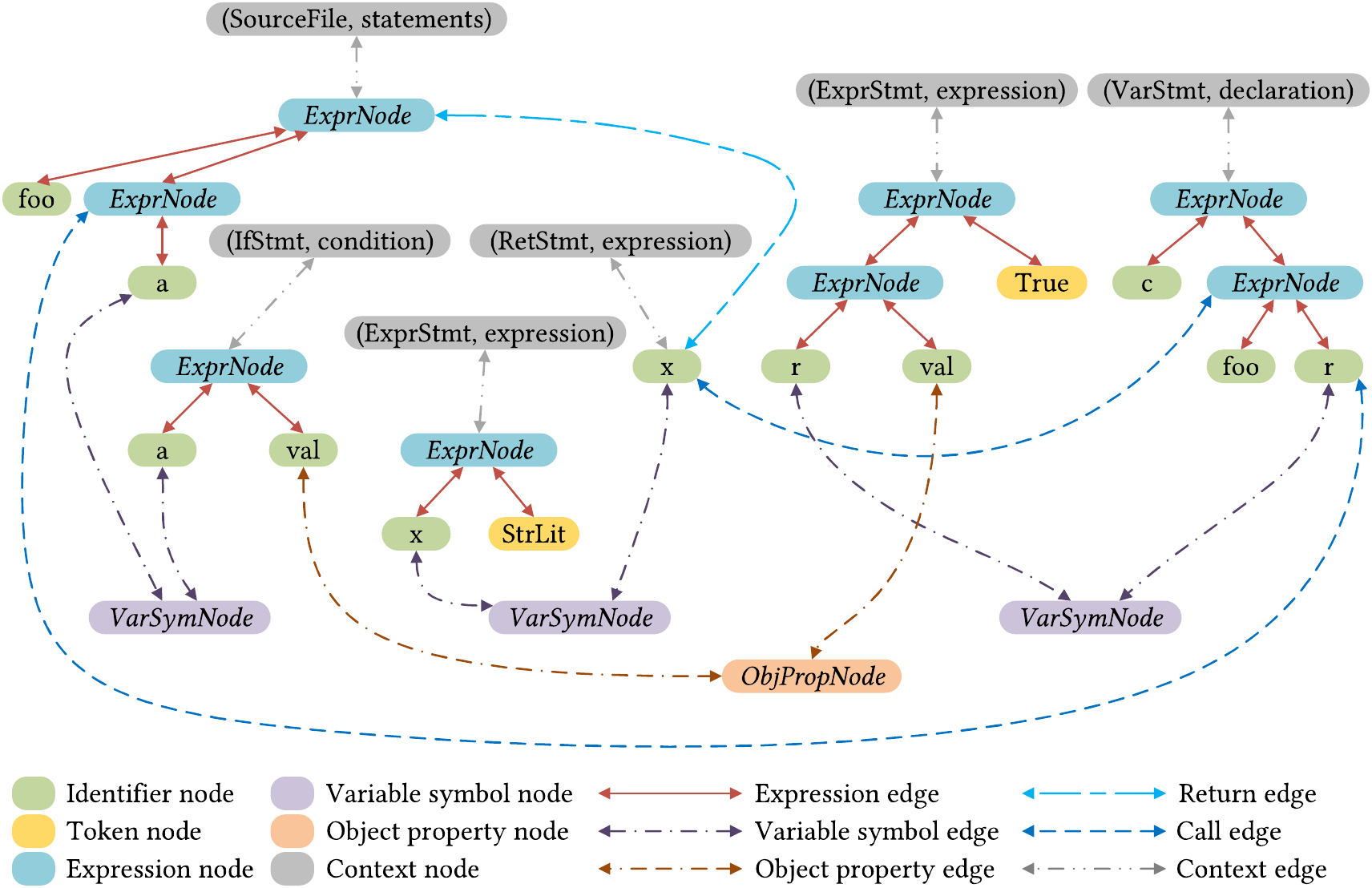}
\caption{Type flow graph.}
\label{fig:tfg_tfg}
\end{subfigure}
\caption{An example of TFG.}
\label{fig:tfg}
\end{figure}

\cref{fig:tfg} gives an example of TFG.
\cref{fig:tfg_code} shows a piece of code and \cref{fig:tfg_ast} displays its AST. The corresponding TFG is shown in \cref{fig:tfg_tfg}.
To simplify the presentation,
the edge features for \expedge{}s are not shown in this example.
It can be seen that \expedge{}s establish type information flow paths within expressions. \varsymedge{}s and \objpropnode{}s 
allow type information exchange between the different occurrences of the same variable and object 
properties.
\retedge{}s allow type information to flow between a function declaration and its return statements.
\calledge{}s enable the information exchange between call sites and their potential callees.%
\footnote{
The call graph construction uses the same name matching mechanism as object properties' 
and thus is also over-approximated.
}
\ctxedge{}s allow expression nodes to get type hints from their context within a statement.

\subsubsection{Graph Extraction}
The graph extraction is based on the traversal of the input program's AST in a bottom-up manner.
For each AST node that matches a TFG node pattern, the graph extractor creates the corresponding 
TFG node and edges based on the definitions described above. For \calledge{} creation, the graph 
extractor runs a lightweight pre-pass to scan the program file to collect function declaration information for later use.

\subsection{Machine Learning Model: Graph Neural Network}
\label{sec:learngnn}
Here we introduce our machine learning model design for static type inference.
This learning model is presented as a customizable system composed of a core GNN model with different building blocks and several other components that converts node/edge features into vector representations.

\subsubsection{Overall Model Architecture}
\label{sec:learnsys}
\begin{figure}[htbp]
    \centering
    \includegraphics[width=0.7\textwidth]{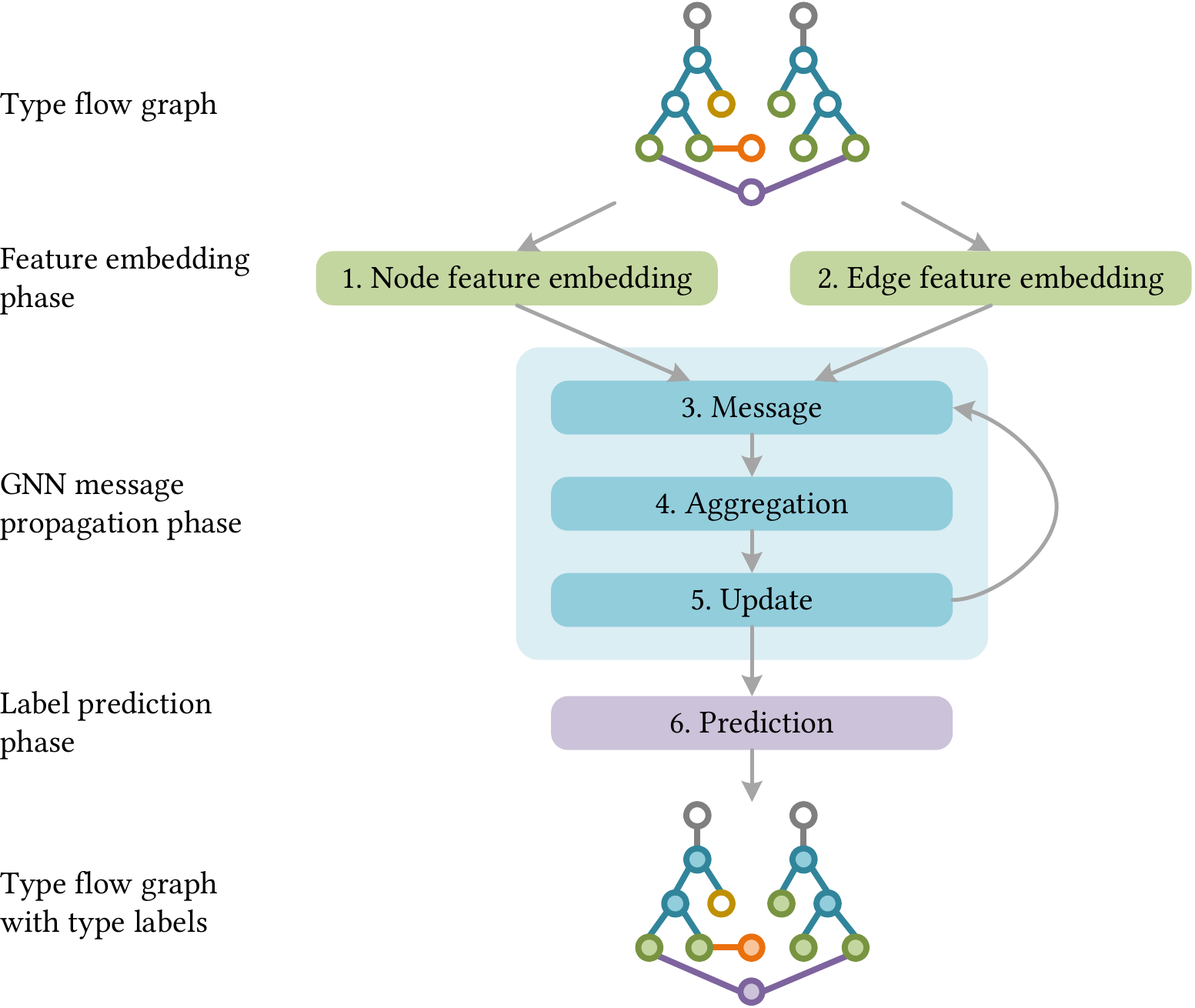}
    \caption{Overall model architecture.}
    \label{fig:gnns}
\end{figure}

As shown in \cref{fig:gnns}, the learning system contains six components that are organized 
into three phases: (1) feature embedding for nodes and edges; (2) GNN message propagation; and (3) label prediction. 

The feature embedding phase (components 1 and 2) translates 
node/edge features in the type flow graph to vector representations (i.e., embedding vectors).
For edge feature embedding and the basic design of node feature embedding (demonstrated in \cref{fig:name_init_base}), we assign a trainable vector to each feature in the vocabulary.
\cref{sec:nsi} discusses several alternative methods for embedding identifier names, which are the features of \identnode{}s.

In the GNN message propagation phase, the node feature embedding vectors are used as node initial states and the edge embedding vectors are used to generate messages on the corresponding edges.
The GNN iteratively propagates and updates node states for $K$ steps using the message, aggregation, and update functions (components 3, 4, 5).
\cref{sec:gnnarch} shows how we define those functions as components of our GNN architecture.

In the label prediction phase (component 6), we use a fully-connected layer followed by a \textit{softmax} function to transform a node's final state after $K$ steps of propagation into a probability distribution over candidate types and output the most probable type as the inferred type for a node.

\subsubsection{\identnode state initialization}
\label{sec:nsi}

Identifier names can sometimes provide hints on the types of associated program elements. For example, a variable named ``\texttt{num}'' usually has an integral type.
We design our model to utilize such type information by treating identifier names as node features of corresponding \identnode{}s and encode them into real-valued embedding vectors, which are used as the initial states of those \identnode{}s.
The GNN then iteratively propagates the type information encoded in the state vectors to other nodes.
\cref{fig:name_init} shows four methods of encoding identifier names that are adopted by our model. \cref{fig:name_init_base} is a basic method that uses a single embedding layer that maps each unique name to a trainable parameter vector. \cref{fig:name_init_seg,fig:name_init_ctx} demonstrate two techniques to improve the name encoding, which will be described in detail in the following paragraphs. The last method shown in \cref{fig:name_init_seg_ctx} is the combination of the two techniques.

\begin{figure}[htbp]
\centering
\begin{subfigure}[t]{\linewidth}
\centering
\captionsetup{justification=centering}
\includegraphics[scale=.6]{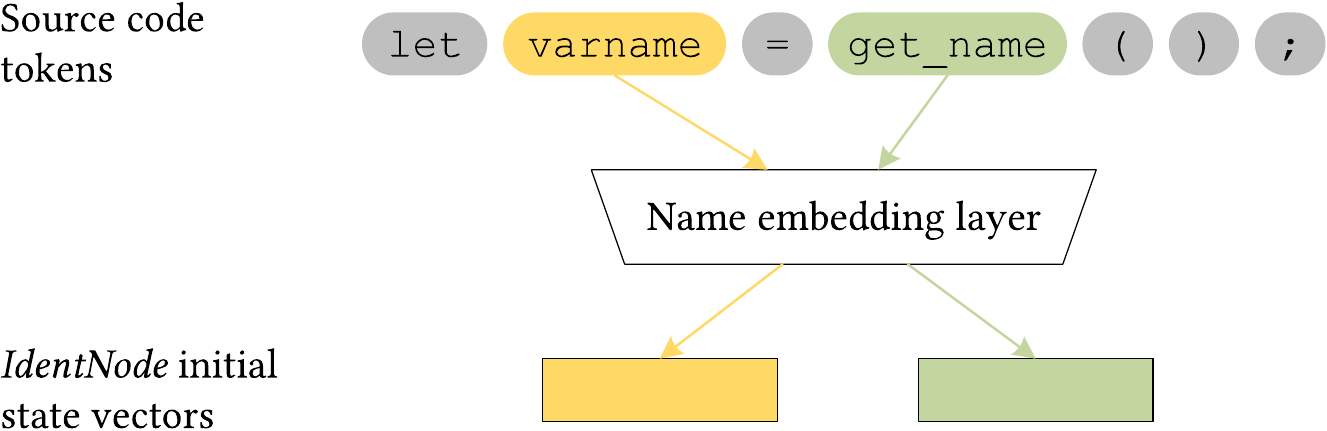}
\caption{}
\label{fig:name_init_base}
\end{subfigure}
\par\vspace{30pt}
\begin{subfigure}[t]{\linewidth}
\centering
\captionsetup{justification=centering}
\includegraphics[scale=.6]{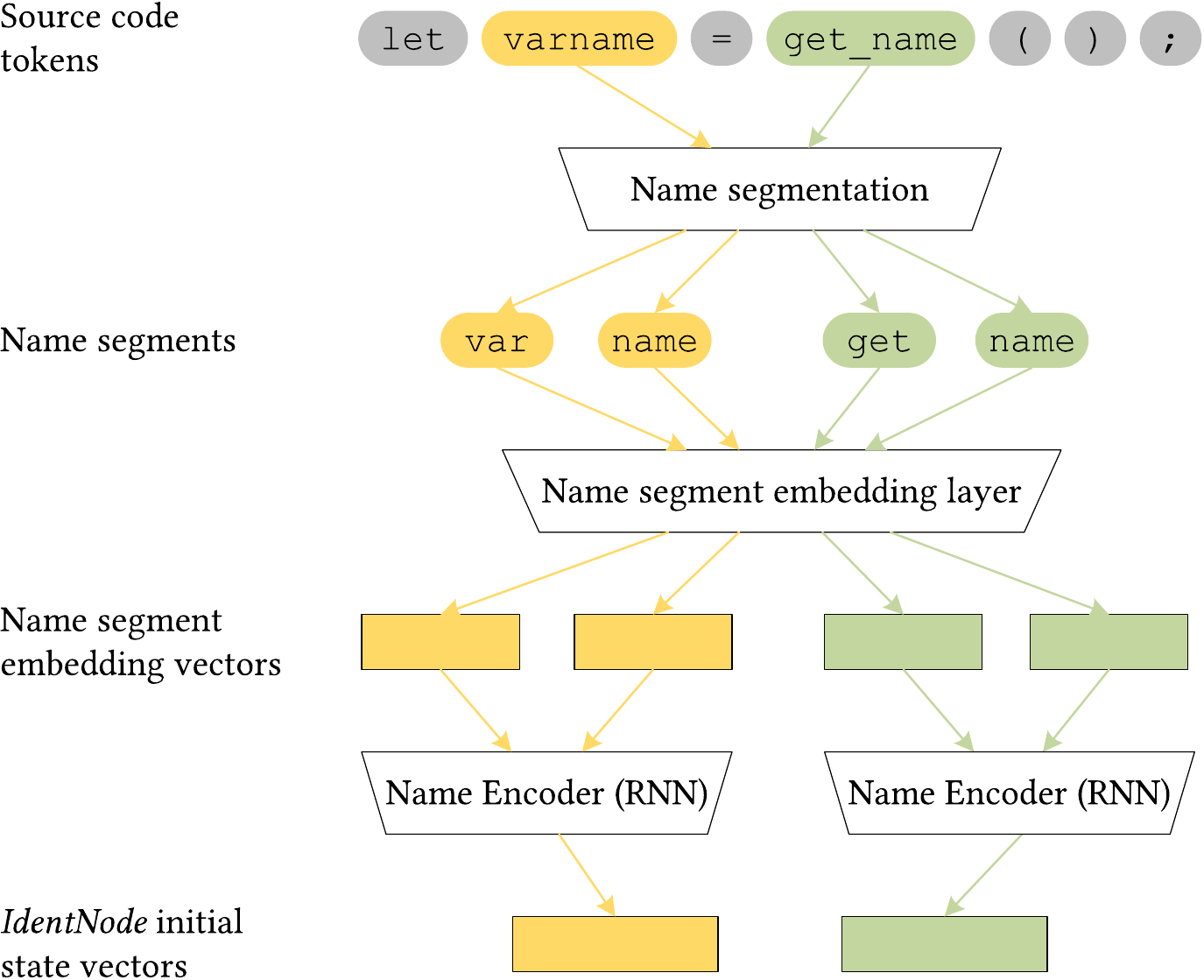}
\caption{}
\label{fig:name_init_seg}
\end{subfigure}
\par\vspace{30pt}
\begin{subfigure}[t]{\linewidth}
\centering
\captionsetup{justification=centering}
\includegraphics[scale=.6]{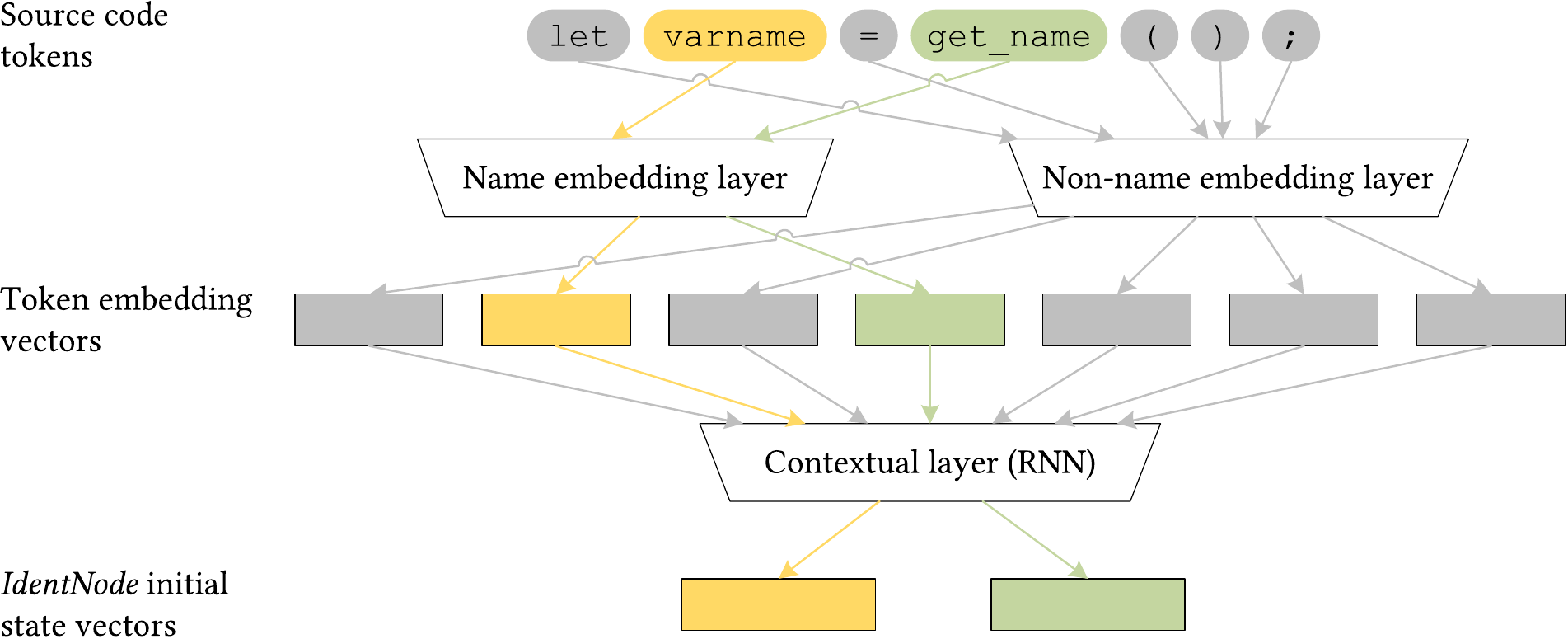}
\caption{}
\label{fig:name_init_ctx}
\end{subfigure}
\end{figure}

\begin{figure}[htbp]
\ContinuedFloat
\centering
\begin{subfigure}[t]{\linewidth}
\centering
\captionsetup{justification=centering}
\includegraphics[scale=.6]{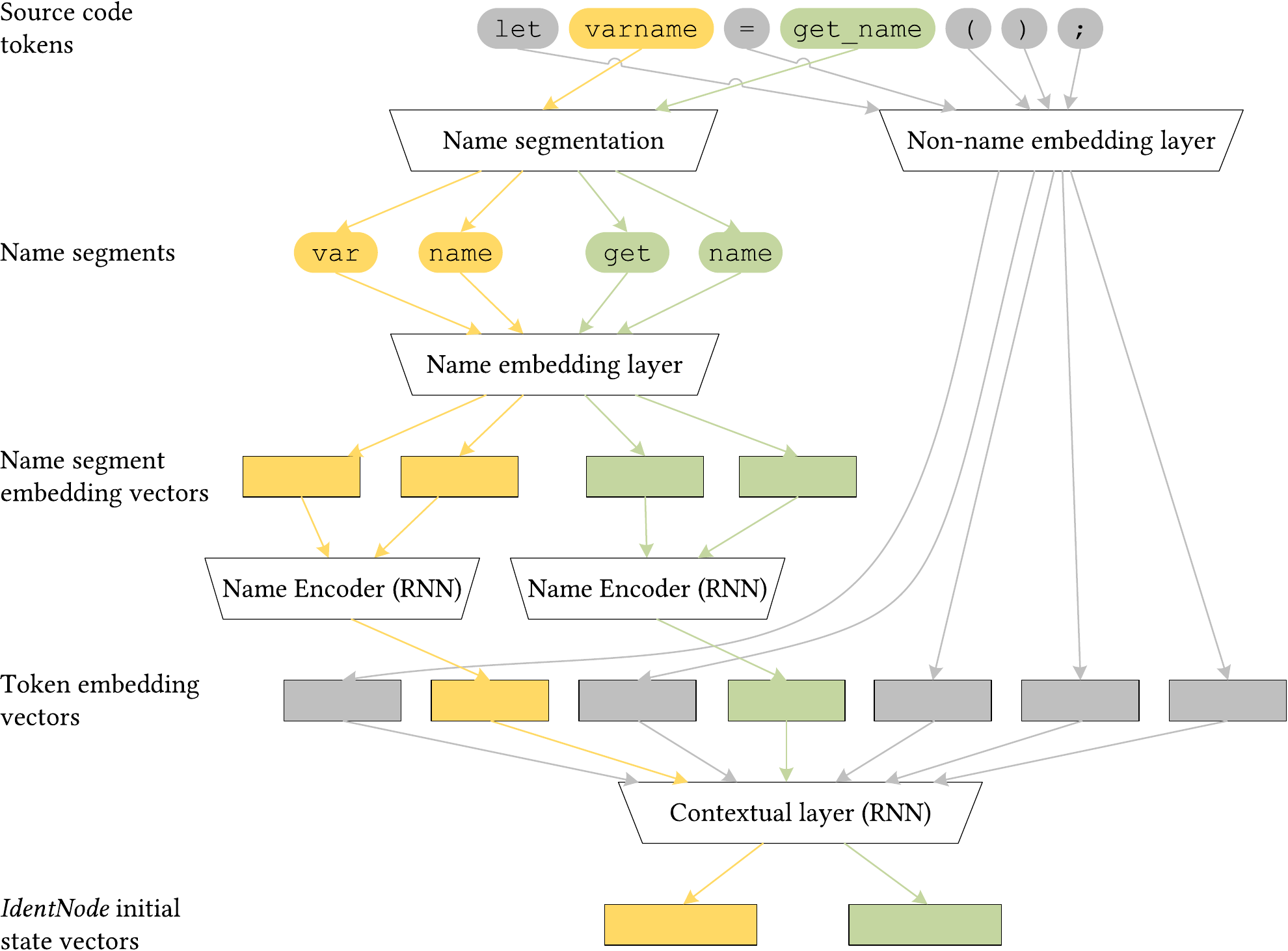}
\caption{}
\label{fig:name_init_seg_ctx}
\end{subfigure}
\caption{\identnode state initialization.}
\label{fig:name_init}
\end{figure}

\paragraph*{Name Segmentation}
The simple embedding method in \cref{fig:name_init_base} requires a fixed-size vocabulary of node identifier names.
However, the size of such vocabulary can be huge due to the diversity of identifier names in programs, making it difficult to learn the information encoded in the names.
Previous works have shown that segmenting identifier names can help reduce the vocabulary size and boost the performance of machine learning models in code modeling~\cite{openvocab,Allamanis17}.
Similar to those prior works, we introduce a {\it name segmentation} mechanism shown in \cref{fig:name_init_seg}. First, an identifier name is split into several segments. Then each of the segments is embedded into a vector through an embedding layer. Finally, a bi-directional RNN is applied to the sequence of segment embedding vectors of an identifier name to generate a single encoding vector, which is used as the initial state vector of the \identnode corresponding to the identifier name.
We perform name segmentation by first splitting an identifier name into subtokens according to the \texttt{CamelCase} and \texttt{snake\_case} patterns, and then further segmenting the subtokens with byte pair encoding~\cite{bpe,nmt}.

\paragraph*{Contextualized Initialization}
The context (surrounding tokens) of an identifier name can also provide useful type information, and thus it could be beneficial to encode it into the initial state of an \identnode.
Although the propagation on the type flow graph can help gather information from the context, the size of such context is limited by the number of GNN propagation steps, which is usually small.
As shown in \cref{fig:name_init_ctx}, to encode the information from a broader range of context for \identnode{}s, we introduce a bi-directional RNN layer that takes the sequence of code token embedding vectors as input and generates an output vector for each token. Those output vectors that correspond to identifiers are used as the initial states of \identnode{}s. We name this RNN layer as the \textit{contextual layer}.

\subsubsection{Graph Neural Network Architecture}
\label{sec:gnnarch}

The \gnn architecture we designed follows the general framework described in \cref{eq:gnn_init,eq:gnn_msg,eq:gnn_agg,eq:gnn_update}.
We describe the instantiation of the framework in the following paragraphs.
Unless otherwise specified, we define the functions below as components of recurrent GNNs and thus do not distinguish between parameters in different propagation steps.

\paragraph*{Message Function}
The functionality of the message function is to produce a message vector $\vec{m}_{uv} \in \mathbb{R}^{d_h}$ that passed from $u$ to its neighbor $v$ that encodes the type information of $u$.
In TFG, we introduce a feature for each edge that is embedded to vector $\vec{e}_{uv} \in \mathbb{R}^{d_e}$.
The edge feature embedding vector is integrated in to the message vector by the following message function:
\begin{flalign}
&& \vec{m}_{uv}^{(k)}
=
\vec{W}_{MO}
\left(
    \left( \vec{W}_{MI} \vec{h}_u^{(k-1)} + \vec{b}_{MI} \right)
    \odot
    \vec{e}_{uv}
\right)
+ \vec{b}_{MO}
&& v \in V, u \in \mathcal{N}(v)
\label{eq:ef_msg}
\end{flalign}
$\vec{W}_{MI} \in \mathbb{R}^{d_e \times d_h}$, $\vec{W}_{MO} \in \mathbb{R}^{d_h \times d_e}$, $\vec{b}_{MI} \in \mathbb{R}^{d_e}$, and $\vec{b}_{MO} \in \mathbb{R}^{d_h}$ are learnable parameters; $\odot$ stands for element-wise multiplication.

Alternatively, we can ignore edge features in the TFG and use the following identity function as the message function:
\begin{flalign}
&& \vec{m}_{uv}^{(k)}
=
\vec{h}_u^{(k-1)}
&& v \in V, u \in \mathcal{N}(v)
\label{eq:id_msg}
\end{flalign}

\paragraph*{Aggregation Function}

After the calculation of message vector $\vec{m}_{uv}$ for all $u \in \mathcal{N}(v)$,
the aggregation function aggregates those message vectors for $v$ into a vector $\vec{a}_{v} \in \mathbb{R}^{d_h}$. A simple solution is to apply a mean aggregation shown below:

\begin{flalign}
&& \vec{a}_{v}^{(k)}
= \frac{1}{|\mathcal{N}(v)|} \sum_{u \in \mathcal{N}(v)} \vec{m}_{uv}^{(k)}
&& v \in V
\label{eq:mean_agg}
\end{flalign}

During type information propagation on the graph, it is likely the neighbors do not pass equally 
important/useful information to a node. For example, the type information coming from a literal 
node may be more informative than that coming from an identifier node.
Consider this simple example:
 \texttt{x = "hello" + y}.
The string literal node \texttt{"hello"} presents stronger information 
than identifier node \texttt{y}.
Thus it could be beneficial to assign different weights to messages from different neighbors during aggregation.

To this end, we introduce an alternative aggregation process that contains an \textit{attention} mechanism to weigh the importance of different incoming messages.
It is adapted from the one in graph attention networks~\cite{gat} and is shown in the following equations:

\begin{flalign}
&& \alpha_{uv}^{(k)}
&= \frac{
    \exp\left(
        \mathrm{LeakyReLU} \left(
            \vec{w}^{\top} \left[ \vec{W}_{QK} \vec{h}_v^{(k-1)} ; \vec{W}_{QK} \vec{m}_{uv}^{(k)} \right]
        \right)
    \right)
}{
    \sum_{u' \in \mathcal{N}(v)}
    \exp\left(
        \mathrm{LeakyReLU} \left(
            \vec{w}^{\top} \left[ \vec{W}_{QK} \vec{h}_v^{(k-1)} ; \vec{W}_{QK} \vec{m}_{u'v}^{(k)} \right]
        \right)
    \right)
}
&& v \in V, u \in \mathcal{N}(v)
\label{eq:attn_weights}\\
&& \vec{a}_{v}^{(k)}
&= \sum_{u \in \mathcal{N}(v)} \alpha_{uv}^{(k)} \vec{W}_V \vec{m}_{uv}^{(k)}
&& v \in V
\label{eq:attn_agg}
\end{flalign}
$\vec{W}_{QK} \in \mathbb{R}^{d_h \times d_h}$, $\vec{W}_{V} \in \mathbb{R}^{d_h \times d_h}$ and $\vec{w} \in \mathbb{R}^{2 d_h}$ are 
parameters to be learned. $\mathrm{LeakyReLU}$ is a non-linear activation function~\cite{relu}.

\paragraph*{Update Function}

In type inference, type information could be propagated many steps across multiple expressions or even procedure boundaries.
To enable effective learning from the potential long-term dependencies, our recurrent GNN architectures integrate the gated recurrent unit (GRU)~\cite{gru} as our update function. This update function has also been adopted in other recurrent GNN architectures~\cite{ggnn}. The update function can be written as:
\begin{flalign}
&&\vec{h}_v^{(k)} = \mathrm{GRU} \left( \vec{a}_v^{(k)}, \vec{h}_v^{(k-1)} \right)
&& v \in V
\end{flalign}

Our GNN architecture can also be configured to be a convolutional GNN, i.e., parameters in functions defined above do not share between propagation steps.
In this case, the update function will not be a recurrent unit, but the one specified below:
\begin{flalign}
&&\vec{h}_v^{(k)} = \mathrm{ReLU} \left( \vec{W}_{h}^{(k)} \vec{h}_v^{(k-1)} + \vec{b}_v^{(k)} \right)
&& v \in V
\label{eq:cgnn_update}
\end{flalign}
$\vec{h}_v^{(k)} \in \mathbb{R}^{d_h \times d_h}$ and $\vec{b}_v^{(k)} \in \mathbb{R}^{d_h}$ are learnable parameters of step $k$.

As an exception, we do not update the states for \toknode{}s and \ctxnode{}s (i.e., their update function is the identity function). This is because the type information they contain is known in advance and thus no information update is needed.

\section{Evaluation}\label{sec:expr}
\newcommand{\rqone}{What are the impacts of our GNN model design choices on the accuracy of type inference?}
\newcommand{\rqtwo}{How efficient is our GNN-based type inference?}
\newcommand{\rqthree}{How does our approach compare with state-of-the-art deep learning type inference approaches?}

In this section, we evaluate our \gnn{} based type inference system to answer the following research questions:
\begin{itemize}
  \item \textit{RQ.1}: \rqone
  \item \textit{RQ.2}: \rqtwo
  \item \textit{RQ.3}: \rqthree
\end{itemize}

\paragraph*{Experimental Platform}
We conducted our experiments on a server that has two Intel Xeon E5-2623 v4 processors with 128GB of RAM and an NVIDIA Tesla V100 GPU with 16GB of graphics memory. The neural network models were trained and evaluated on the GPU.

\paragraph*{Dataset and Preprocessing}
We constructed our dataset based on a corpus consisting of popular
TypeScript projects in Github that were also used in past work on type
inference for JavaScript~\cite{deeptyper}.
We obtained the projects that are still publicly available from that
set, and, following the methodology in \cite{deeptyper}, removed all
files containing more than 5,000 tokens so as to enable efficient batching.
The resulting set of projects was then randomly split into {\em training}, {\em
  validation} and {\em test} subsets, which respectively contain 789, 99 and 99 projects,
or 74,801, 3,729 and 3,838 files.

\begin{figure}[h]
  \centering
  \includegraphics[width=0.6\textwidth]{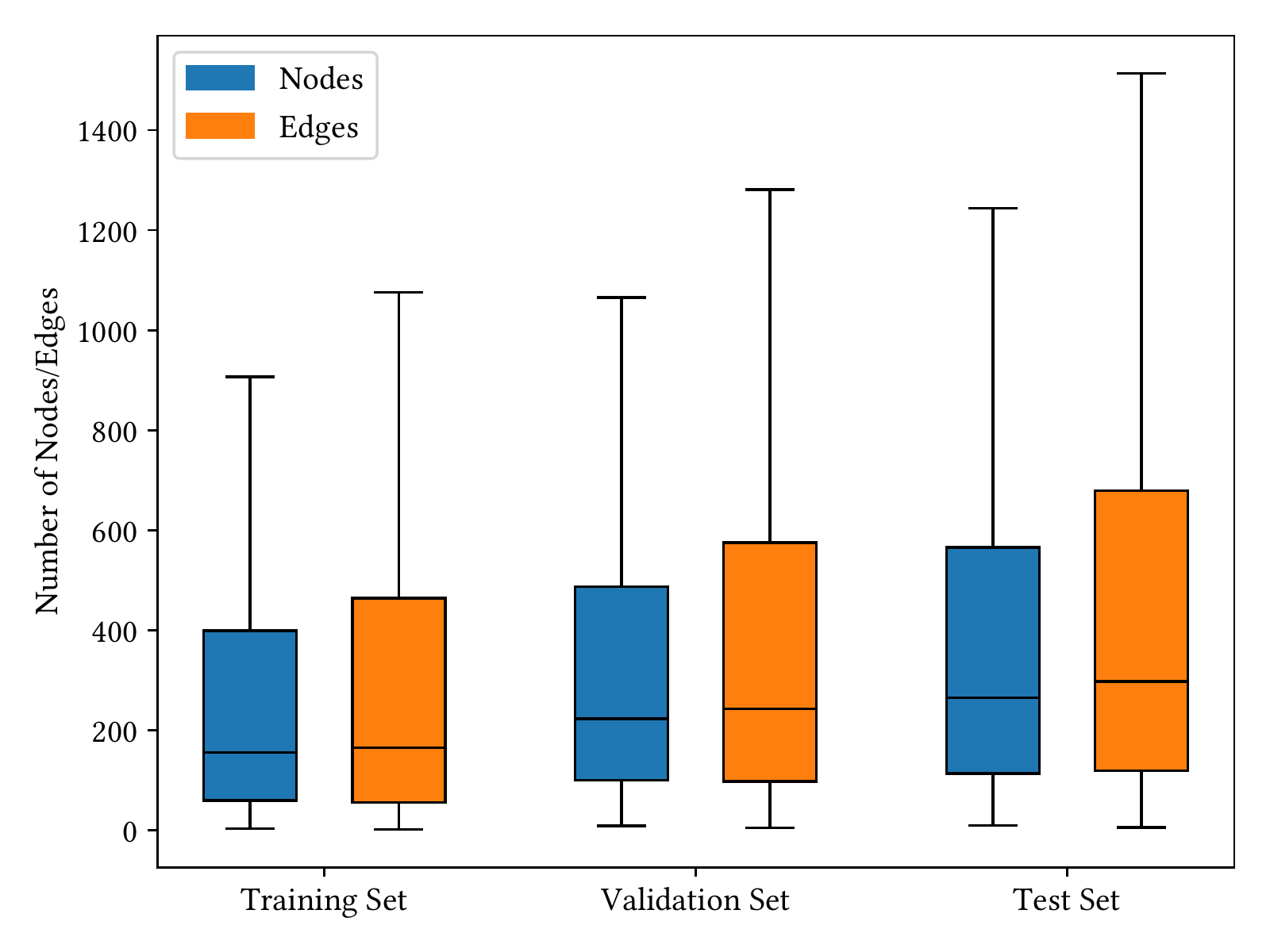}
  \caption{Distributions of the numbers of nodes/edges in a graph for
    the training, validation, and test sets which contain 74,801,
    3,729 and 3,838 files/graphs respectively, shown as
    box-and-whiskers plots.  Each box spans values from the first
    quartile ($Q1$) to the third quartile ($Q3$), with the median shown as a
    horizontal line in the box.  The endpoint whiskers indicate the
    values, $Q1 - 1.5 \times(Q3-Q1)$ and $Q3 + 1.5 \times (Q3-Q1)$, for the respective datasets.}
  \label{fig:graphsize}
\end{figure}

Our toolchain was used to
construct a type flow graph (TFG) for each file.
\cref{fig:graphsize} shows the distributions of the TFG size in the
training, validation, and test sets, in terms of  numbers of nodes and
edges.
Note that TypeScript type annotations are not used when constructing
the TFGs.

To identify type labels in the training set, we used the TypeScript compiler
to extract the types for all nodes in our TFGs, including
identifier nodes and other expression nodes.  Since TypeScript permits
partial type annotations, these type labels can include types provided
by the developer as well as types inferred by the TypeScript compiler.
The extracted types were then preprocessed with the following steps:
(1) function types are mapped to their return types;
(2) the type parameters in parametric types are removed (e.g., \texttt{Array<number>} becomes \texttt{Array}); 
(3) literal types are mapped to their base types (e.g., the literal type \texttt{"a"|"b"} becomes \texttt{string});
(4) types with names that consist of a single character are filtered out, as they usually represent type parameters (e.g., \texttt{T}, \texttt{U}, \texttt{V}).
A type vocabulary was built after the preprocessing, and the models were trained to predict types from this vocabulary.
Similar to \cite{lambdanet}, we picked the top-100 frequent types from the training set, excluding the \texttt{any} type.
We did not use a larger type vocabulary because our focus is on
predicting types that are used across different projects rather than
types defined locally in a project. Prediction of locally defined
types is beyond the scope of this paper/

We constructed fixed-sized vocabularies for graph edge features, non-name node features, and identifier names (or name segments) for them to be embedded as trainable vectors in the neural 
networks. We included all edge features in the training set in the edge vocabulary. The 
resulting edge feature vocabulary has a size of 210. We also constructed a vocabulary of non-name features in the same manner and its size is 184. For names, we constructed a 
vocabulary of full names when we do not perform name segmentation, and a vocabulary of name 
segments otherwise. The full name vocabulary contains the 10,000 most frequent identifier 
names in the training set.
We also include a special \texttt{UNKNOWN} token to represent out-of-vocabulary names.
For name segmentation, we generated the segment vocabulary by 
performing up to 10,000 merges during byte pair encoding on the training set.

When performing inference on files in the validation and test sets, we followed the methodology 
adopted by prior work~\cite{deeptyper,lambdanet}, and  only used labels available in type 
annotations provided by the developer (and ignored the types inferred by the TypeScript
compiler). Doing so only assigns labels to identifier nodes. We then repeat steps (1) to (4) 
listed above for these extracted types in the validation and test sets. We also removed edges 
with out-of-vocabulary edge features from the type flow graphs built from the validation and 
test sets.

\paragraph*{Evaluated Model Designs}
As mentioned in Section~\ref{sec:gnnarch}, we studied different
\gnn{} model design choices, including the setup for node state initialization and the \gnn{}
architecture.  The motivation for evaluating different model designs
is to gain insight as to which is the best learning system for type
inference, and to answer \textit{RQ.1}: ``\rqone{}''

Table~\ref{tab:networkdef} lists the various \gnn designs (along with their
different design options) that are encompassed by our evaluation.
\gggn is our baseline model, which is a recurrent \gnn (i.e., a \gnn that shares 
parameters across propagation steps) that uses the \gru update function and mean message aggregation.
It takes in edge features but does not perform name segmentation or contextualized initialization.
\cgnn is a variant of \gggn that does not share parameters across steps and uses the update function defined in \cref{eq:cgnn_update}.
\ggat extends \gggn with the attention-based message aggregation function defined in \cref{eq:attn_weights,eq:attn_agg}.
\gggnns introduces name segmentation in the node initial state generation phase to \gggn.
Similarly,  \gggnctx introduces the contextual layer to \gggn.
\gggnnsctx adds both name segmentation and the contextual layer into
\gggn.  We observe that none of these eight \gnn{} variants have been
studied before in past work on type inference.

All of the  \gnn{} designs described above are based on the TFG with edge features.
To evaluate the impact of not including edge features in TFGs, and the effectiveness 
of the attention mechanism in such situations, we built two more models that do not 
take edge features as input. They are \gggnnef and \ggatnef, which correspond to  
\gggn and \ggat, respectively.

\begin{table*}[htbp]
\centering
\caption{Designs of \gnn{} Architectures.}
\label{tab:networkdef}
\resizebox{\columnwidth}{!}{%
\begin{tabular}{l|c|c|c|c|c}
 & \gnn{} Type & Attention & Name Segmentation & Contextual Layer & Edge Features \\\hline 
 \cgnn & Convolutional & & & & $\checkmark$ \\
 \gggn & Recurrent & & & & $\checkmark$ \\
 \ggat & Recurrent & $\checkmark$ & & & $\checkmark$ \\
 \gggnns & Recurrent & & $\checkmark$ & & $\checkmark$ \\
 \gggnctx & Recurrent & & & $\checkmark$ & $\checkmark$ \\
 \gggnnsctx & Recurrent & & $\checkmark$ & $\checkmark$ & $\checkmark$ \\
 \gggnnef & Recurrent & & & & \\
 \ggatnef & Recurrent & $\checkmark$ & & & \\
\end{tabular}%
}
\end{table*}
\paragraph*{Evaluation Metrics}
We evaluated the models in terms of their type prediction accuracy and efficiency.
The accuracy metrics include top-1 and top-5 accuracy for predicting types for the test set.
Those two kinds of accuracy metrics are obtained by using the model to
output the top-1 and top-5 probable types for each prediction location
and then computing how often  the ground truth type label matches the output type(s).
We divided types in the vocabulary into two categories: the 10 most frequent types in the 
training set and the other 90 types, and measured the top-1 and top-5 accuracy for all 100 
types and the two categories separately. The efficiency metrics we used are the number of 
parameters in the model and its throughput (files/sec) for inference.

\paragraph*{Implementation Details}
We used the TypeScript compiler to extract type labels and generate ASTs 
for source code files in the dataset. Our TFGs were constructed from those ASTs.
The neural network architectures used in our experiments were built using PyTorch~\cite{pytorch}.
In all experiments, we fixed the batch size to be 64 and trained the models for 60 epochs using the 
AdamW optimizer~\cite{adamw} with a learning rate of $10^{-3}$. We evaluated the model on the validation
 set after each epoch and used the model from the epoch that produced the lowest validation loss to be 
evaluated on the test set. We used 128-dimensional node type/name embeddings, 32-dimensional name 
segment embeddings, 32-dimensional RNN hidden states for name segment sequence encoding, 256-dimensional
 edge type embeddings, 128-dimensional node states, and 128-dimensional RNN hidden states for both 
directions in the contextual layer.
For the comparison with \deeptyper{} and \lambdanet{}, we implemented their models in our evaluation framework and used the same hyperparameters as suggested in their papers and open-source implementations.

\subsection{\textit{RQ.1}: \rqone}
\label{sec:rq1}
As discussed in Section~\ref{sec:gnn}, our GNN-based type inference system allows multiple 
design choices across the different components. We studied how they could affect the accuracy 
of type prediction by evaluating various combinations of those design choices on our dataset.

\subsubsection{The Number of GNN Propagation Steps}
\begin{figure}[h]
\centering
\includegraphics[width=0.8\textwidth]{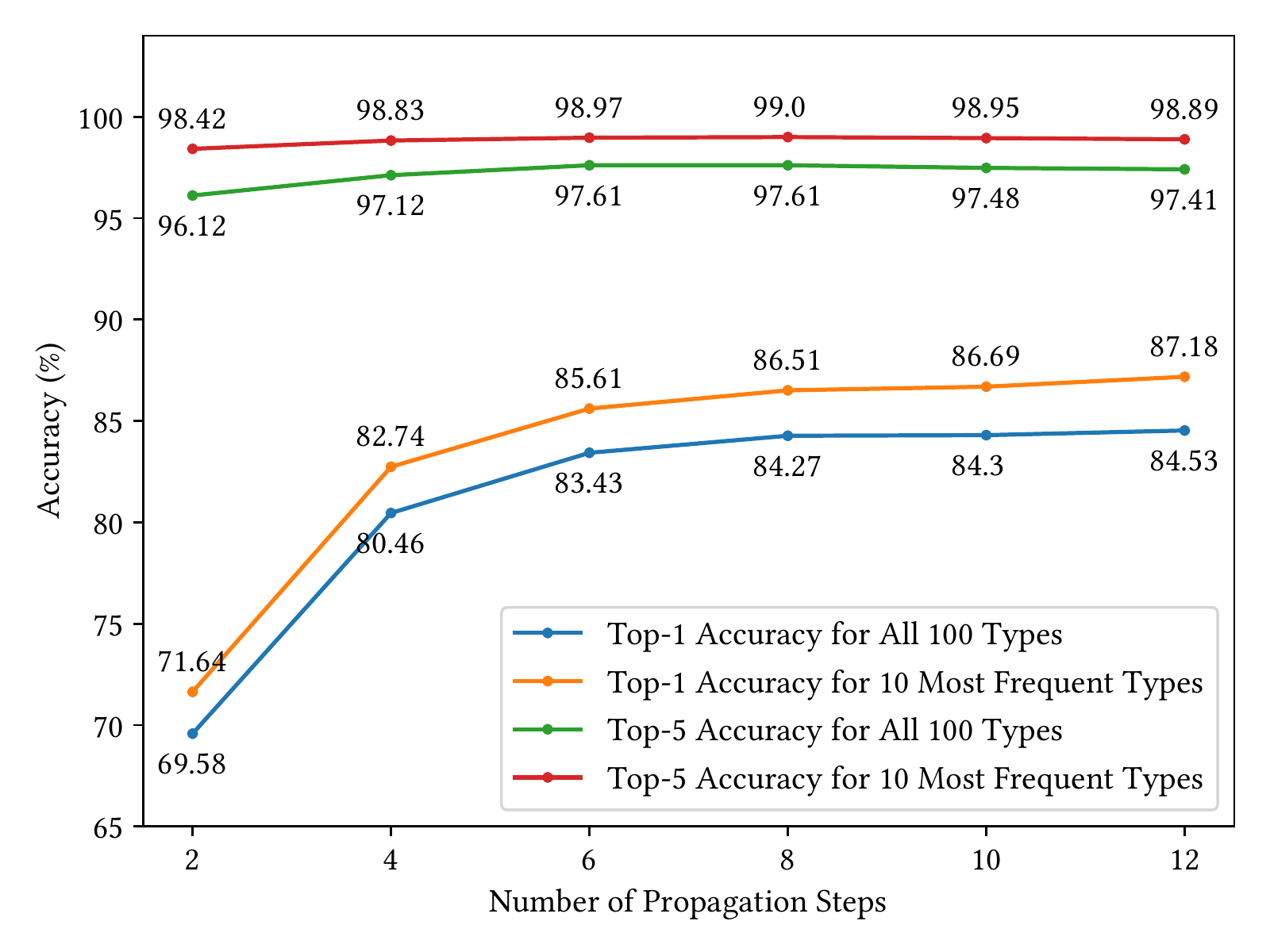}
\caption{The impact of the number of propagation steps ($K$) on type prediction accuracy for the \gggn model.}
\label{fig:prop_steps}
\end{figure}

The  \gnn{} architecture includes a $K$-step iterative process of propagating information on the input graph.
This means a node can gather type information from nodes that are at most $K$ hops away. Intuitively, a 
larger value of $K$ would likely lead to a higher type prediction accuracy since more information can be 
utilized for making the prediction. However, as the value of $K$ increases, the model needs more
computation resources. 
To select an appropriate value of $K$, we made an empirical study: trained the \gggn model with values 
of $K$ ranging from 2 to 12 and measured the resulting accuracy on the test set. The results are shown 
in \cref{fig:prop_steps}. As can be seen from the figure, the top-1 accuracy increases monotonically as 
the value of $K$ gets larger, but begins to saturate around the point where $K=8$.
Based on this observation, we chose $K=8$ for the rest of our experiments.

\subsubsection{Recurrent GNN vs. Convolutional GNN}
A recurrent GNN architecture uses the same set of functions (message, aggregation and update functions) 
and parameters for different steps to aggregate messages from neighbors and update node states, while a 
convolutional GNN does not share parameters or even functions across steps.
As a result, recurrent GNNs can learn to propagate the same type of information between the nodes through 
several time-steps, and convolutional GNNs usually learn a different kind of node representation at each step.
We hypothesize that recurrent GNN architectures can fit the type inference problem better because it can potentially model type propagation rules on TFGs.
To validate this hypothesis, we compared the performance of two architectures: \gggn, our baseline recurrent GNN, and its convolutional variant, \cgnn.
As shown in Part 1 of \cref{tab:comp_precision}, \gggn outperforms
\cgnn in all of the six accuracy metrics, indicating that our hypothesis
holds true for the datasets used in our evaluation.  However, since
the difference in the accuracy metrics is small in some cases, further
study may be warranted to better understand the
capability of both architectures on type inference as part of future work.

\begin{table*}[htbp]
\begin{center}
\caption{The accuracy comparison among various \gnn{} designs and prior work.}
\label{tab:comp_precision}
\begin{tabular}{l|c|c|c|c|c|c}
\multicolumn{1}{c|}{} & \multicolumn{2}{c|}{All 100 Types} & \multicolumn{2}{c|}{Top-10 Frequent Types} & \multicolumn{2}{c}{Other 90 Types} \\
Accuracy & {\it Top-1} & {\it Top-5}  & {\it Top-1} & {\it Top-5} & {\it Top-1} & {\it Top-5} \\\hline

\multicolumn{7}{l}{\multirow{2}{*}{{\it Part 1: recurrent GNN vs. convolutional GNN}}} \\
\multicolumn{7}{l}{\multirow{2}{*}{}} \\\hline

\cgnn              & 83.10\%    & 96.99\%   & 85.70\%   & 98.68\%  & 52.33\%      & 77.04\%    \\
\gggn              & 84.27\%   & 97.61\%   & 86.51\%  & 99.00\%   & 57.77\%      & 81.23\%    \\\hline

\multicolumn{7}{l}{\multirow{2}{*}{{\it Part 2: the impact of the attention mechanism and edge features}}} \\
\multicolumn{7}{l}{\multirow{2}{*}{}} \\\hline

\ggat              & 83.62\%   & 97.37\%   & 86.22\%  & 98.94\%  & 52.91\%      & 78.84\%    \\
\gggnnef           & 79.91\%   & 97.17\%   & 82.37\%  & 99.96\%  & 50.82\%      & 75.99\%    \\
\ggatnef           & 80.45\%   & 97.08\%   & 82.83\%  & 98.80\%   & 52.37\%      & 76.83\%    \\\hline

\multicolumn{7}{l}{\multirow{2}{*}{{\it Part 3: the impact of name segmentation and the contextual layer}}} \\
\multicolumn{7}{l}{\multirow{2}{*}{}} \\\hline

\gggnns            & 86.89\%   & 97.66\%   & 89.43\%  & 98.80\%    & 56.81\%      & 84.25\%    \\
\gggnctx           & 86.43\%   & 98.00\%    & 88.29\%  & 99.08\%   & 64.43\%      & 85.21\%    \\
\gggnnsctx         & 87.76\%   & 97.99\%   & 90.31\%  & 99.18\%   & 57.56\%      & 83.91\%    \\\hline

\multicolumn{7}{l}{\multirow{2}{*}{{\it Part 4: state-of-the-art deep learning type inference}}} \\
\multicolumn{7}{l}{\multirow{2}{*}{}} \\\hline

{\it DeepTyper}    & 84.62\%   & 97.79\%   & 86.88\%  & 99.08\%   & 57.94\%      & 82.57\%    \\
{\it LambdaNet}    & 79.45\%   & 96.83\%   & 82.32\%  & 98.61\%   & 41.84\%      & 73.52\%    \\
\end{tabular}
\end{center}
\end{table*}

\subsubsection{The Attention Mechanism and Edge Features}
The attention mechanism in message aggregation allows for assigning different importances to messages received from the neighborhood of a node,
and thus can help filter out noise or irrelevant information coming from the neighbors.
Here we study its effectiveness when applied to our type inference framework.
Based on the results shown in Table~\ref{tab:comp_precision} parts 1 and 2, the precision of \gggn, which applies a mean aggregation (\cref{eq:mean_agg}), is better than \ggat, which applies an attention-based aggregation (\cref{eq:attn_weights,eq:attn_agg}), across all accuracy metrics.
This implies that attention did not result in a benefit for our framework on the inference problem studied in our evaluation.

As the messages are the results of combining neighbors' node states and edge features (see \cref{eq:ef_msg}), those edge features may help encode enough information of a message's importance into it, and the usefulness of an attention mechanism could be reduced due to this reason.
To figure out the impact of edge features and their interaction with the attention mechanism, we also compared the accuracy of models that do not take edge features as input.
As shown in part 2 of \cref{tab:comp_precision}, the two models,  \gggnnef 
and \ggatnef, which correspond to \gggn and \ggat, but with no edge
feature inputs, yield worse accuracy than their counterparts,
indicating that edge features play an important role in the message aggregation process.
However, \ggatnef still fails to show significant advantages over \gggnnef.
This indicates that the attention mechanism may not be able to show its effectiveness on our TFG structure, no matter whether edge features are present or not.

\subsubsection{\identnode State Initialization}
There is a general belief that identifier names convey programmers' intuition related to types.
The \identnode state initialization techniques discussed in Section~\ref{sec:nsi} explore the potential of using identifier names and their context to help with type prediction.
There are two improvements in node state initialization relative to the baseline method used in \gggn, which simply assigns a trainable embedding vector to a name and uses it as the initial state for the corresponding identifier nodes.
\gggnns incorporated the first improvement -- name segmentation -- into \gggn, and \gggnctx added the second improvement -- the contextual layer -- to \gggn.
\gggnnsctx combined the two improvements together.
In part 3 of \cref{tab:comp_precision}, we compared
the accuracy among the three improved models with \gggn.
All of the three models produced higher accuracy than \gggn,
and the model that combines the two techniques (\gggnnsctx) gave better accuracy than the two models that only incorporated one of the techniques (\gggnns and \gggnctx).
This indicates that the combination of name segmentation and the contextual layer can effectively capture type-relevant information and produce high-quality name embeddings to serve as \identnode initial states, which can benefit GNN-based type prediction.

\subsection{\textit{RQ.2}: \rqtwo}
\label{sec:rq2}

\begin{figure}[htbp]
\centering
\begin{subfigure}[t]{0.49\linewidth}
\includegraphics[width=\linewidth]{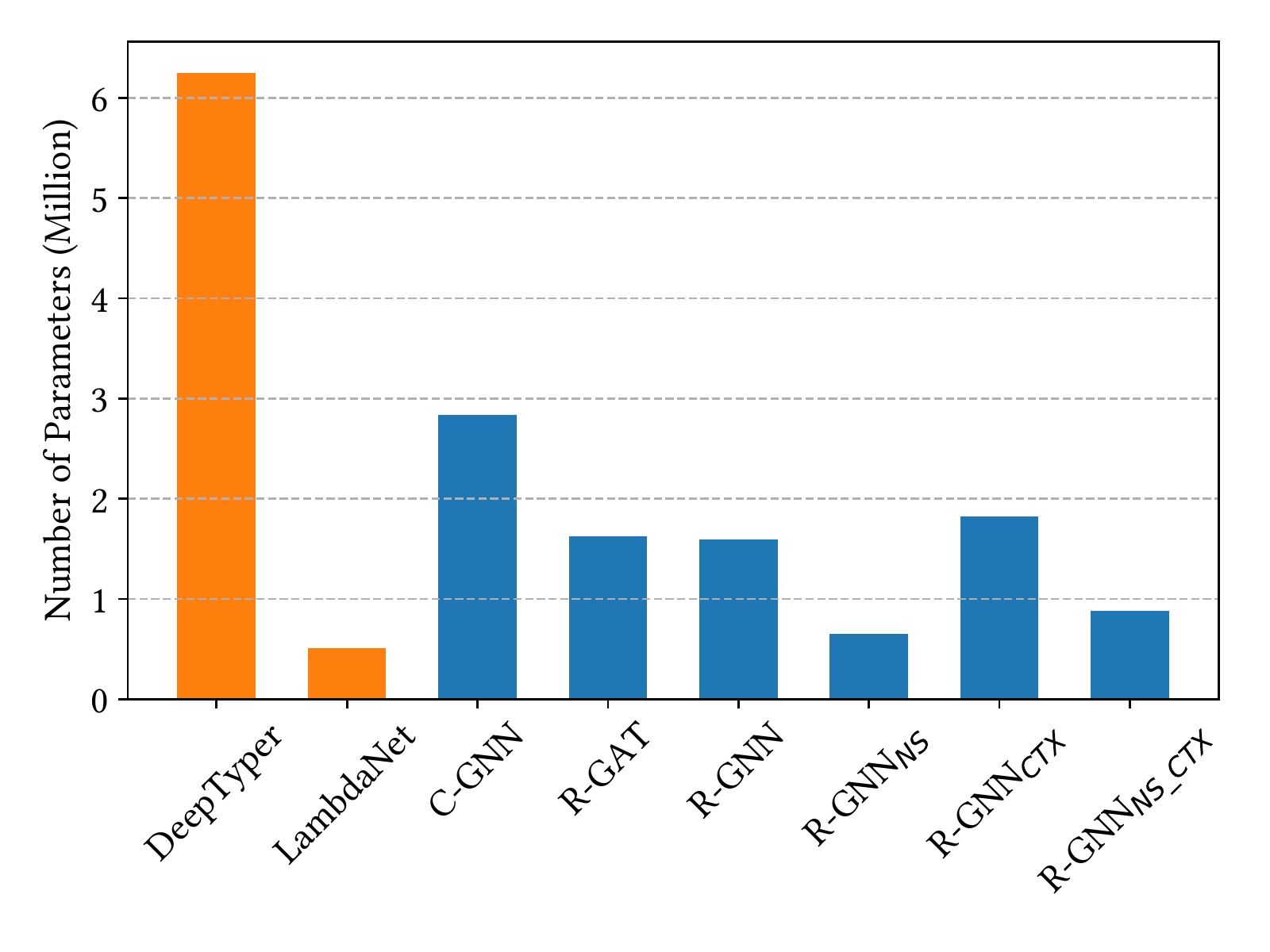}
\caption{Number of parameters in a model.}
\label{fig:param}
\end{subfigure}
\begin{subfigure}[t]{0.49\linewidth}
\includegraphics[width=\linewidth]{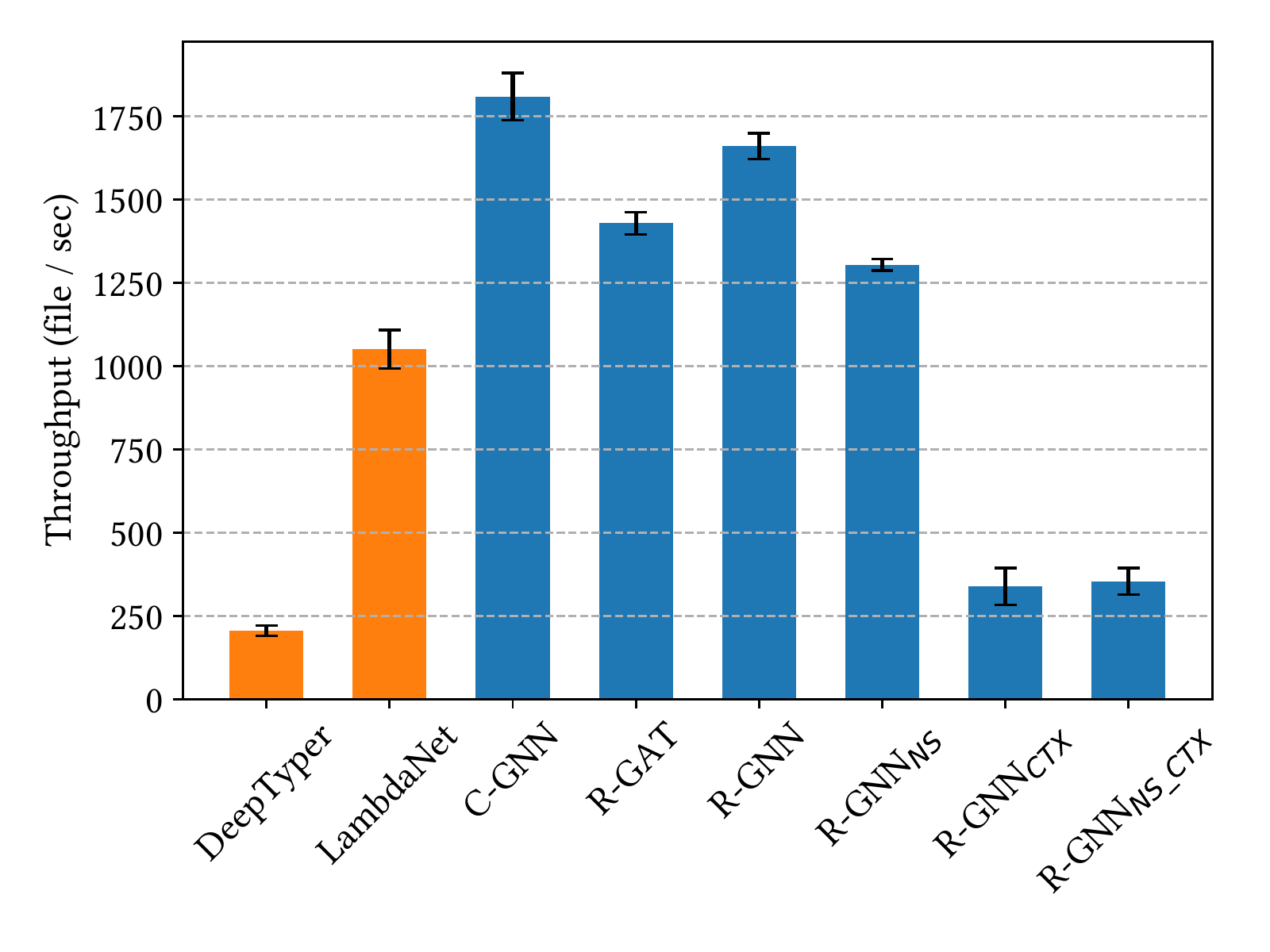}
\caption{Inference throughput on the test set. The error bars show the standard deviation of 6 measurements.}
\label{fig:throughput}
\end{subfigure}
\caption{The comparison of the model size and computation efficiency among our GNN architectures and models proposed in prior work.}
\label{fig:efficiency}
\end{figure}

For our approach to be applicable to real-world interactive programming
environments, it is important for our techniques to be efficient in
practice.
To evaluate the efficiency of the models in our framework, We measured their sizes 
(i.e., the number of model parameters) and inference throughputs.
The results are shown in \cref{fig:efficiency}. Here we only compare the models 
that take in edge features.

The number of parameters in a model affects its applicability in several ways. A 
larger model costs more memory to be loaded and can sometimes bring more computation workload. 
The blue bars in \cref{fig:param} shows the number of parameters of our GNN models.
\cgnn contains the largest number of parameters due to its non-recurrent design 
that requires a separate set of parameters for each step. The size of \ggat is 
slightly larger than \gggn because of its attention mechanism that requires extra parameters.
Among the \gggn variants, the two with name segmentation are significantly smaller 
than the others thanks to their smaller name segment embedding size (32-dimensional), 
although the contextual layer increases the model size by bringing in more parameters.

\cref{fig:throughput} compares the computation efficiency of the models by showing their inference throughputs.
This metric is computed as the number of source files in the test set divided by the 
total runtime of performing inference on those files.
The batch size used for inference is set to 64 graphs for all models.
We took six measurements for each model and reported their mean and standard deviation.
\cgnn and \gggn are the two simplest models in terms of the architecture, requiring less computation 
during inference, and thus produced the highest throughputs.
The attention mechanism in \ggat adds additional computation overhead, causing it to produce a lower throughput than \gggn.
Similarly, name segmentation and the contextual layer both make the models containing them to be slower.
It is worth noting that the addition of the contextual layer brings more computation overhead. 
This is because it is based on a recurrent neural network, whose workload is determined by the length of the input code token sequence.

\subsection{\textit{RQ.3}: \rqthree}
\label{sec:rq3}
We compared our approach with two state-of-the-art deep learning type inference approaches:
\deeptyper~\cite{deeptyper} and \lambdanet~\cite{lambdanet}.

\deeptyper{} treats a source file as a sequence of tokens and runs a two-layer bidirectional RNN 
on it to predict types for identifier tokens. To improve the consistency of type prediction for 
the same variable, the model also includes a consistency layer between the two RNN layers that 
takes the mean of hidden states for identifiers with the same name and adds it back to those hidden states.
We implemented \deeptyper{} in our evaluation framework and used the same hyperparameters as suggested 
in their paper and open-source implementation (300-dimensional embeddings and 650-dimensional hidden layers).%
\footnote{https://github.com/DeepTyper/DeepTyper}

\lambdanet{} is a GNN-based neural type inference model.
It extracts a project-level type dependency graph (TDG) through static analysis and uses a convolutional GNN to perform type inference for the graph nodes,
each of which is associated with a unique variable or an expression.
\REV{%
\lambdanet{}'s TDG includes a manually defined set of edge types and labels. In contrast, our TFG automatically derives hundreds of edge labels from AST node types and parent-child relationships, which provide richer information to the GNN model and is also more easily adaptive to other languages.%
}
We also implemented \lambdanet{} in our evaluation framework.
Unlike the original \lambdanet{}, our implementation gives a prediction for each occurrence of a variable instead of having one prediction per declaration, and uses a prediction space consisting of types in a fixed vocabulary.
\REV{%
Those changes allow us to compare the efficacy of \lambdanet{}'s TDG and our TFG without the potential interference brought by different prediction schemes.%
}
We didn't change the neural network architecture and used the hyperparameters suggested in their paper and open-source implementation (32-dimensional 
embeddings and hidden layers, six steps of graph propagation).%
\footnote{https://github.com/MrVPlusOne/LambdaNet}

\REV{%
The reason why we chose to reimplement the two state-of-the-art approaches in our own framework, instead of using their released code, is to obtain an apples-to-apples comparison. We studied both \deeptyper{}'s and \lambdanet{}'s code repositories and had the following observations:
\begin{itemize}
    \item Their training code is tightly coupled with their data processing code, making it hard to reuse their code with a different data preprocessing approach such as that in our approach.
    \item Their implementations are based on different deep learning frameworks which could result in efficiency variations for the same model, bringing challenges to our efficiency comparison.
    \item \lambdanet{}'s authors implemented a variant of \deeptyper{} in their framework for comparison. However, our preliminary experiments showed that some changes introduced by the variant had a negative impact on accuracy on our dataset.
\end{itemize}
Based on those observations, we decided to use our own implementations of the two approaches in the experiments.%
}

Part 4 of \cref{tab:comp_precision} shows the accuracy \deeptyper{} and \lambdanet{} achieved in our 
experiments. 
\deeptyper{} delivered a similar level of accuracy 
compared with \gggn, our baseline GNN model. However, it is outperformed by the three variants 
of \gggn that integrates name segmentation and/or the contextualized layer in terms of top-1 
accuracy metrics. \lambdanet{} failed to reach the same level of accuracy as \gggn. In terms of 
top-1 accuracy for predicting all 100 types, our best model, \gggnnsctx, outperformed \deeptyper{} 
by 3.14\% (absolute) and \lambdanet{} by 8.31\%. Note that since we ported both prior approaches 
into our experimental framework, which contains different data processing procedures, the accuracy 
numbers we obtained could not be directly compared with the ones reported in their papers.
This is especially the case for \lambdanet due to all the changes we made to it in our implementation.

The orange bars in \cref{fig:efficiency} shows the sizes and inference throughputs of \deeptyper and 
\lambdanet. As can be seen from \cref{fig:param}, \deeptyper contains significantly more parameters 
than other models. This is due to its large hidden state size. In contrast, \lambdanet employs a 
relatively small hidden state size, resulting in a more compact model. 
The sizes of our GNN models fall in between and are closer to \lambdanet's. Shown in 
\cref{fig:throughput}, 
\deeptyper yielded the lowest throughput among all the models being compared.
This is because \deeptyper runs multiple layers of RNN through the code token sequence, which can be 
very long. Our models with the contextual layer (\gggnctx and \gggnnsctx) suffered from the same situation, 
but still gave higher throughputs than \deeptyper, 
because that the contextual layer used only a single layer of RNN. The other four \gnn{} models without 
the contextual layer achieved higher throughputs than \lambdanet.
The reason comes from that \lambdanet generally contain more edges than our TFGs, thus requiring 
more computation resource during message passing and aggregation.

\section{Related Work} \label{sec:relwork}
As mentioned earlier, there has been a significant amount of work
during the past two decades on applying machine learning to address a
number of compilation-related problems, and this work has been enabled by the rapid growth of open source online code repositories.  We only discuss the most closely related efforts from past work in this section.

JSNice~\cite{jsnice} is a program property predictor for variable names and type 
information. It applies probabilistic reasoning to infer primitive types in JavaScript code. Its 
input is a dependency network among variables in JavaScript. The trained model learns the statistical 
correlations among the node in the networks extracted from the code corpus, and predicts type information for 
variables by giving the rank of probabilities.
More recently, a deep neural network based type inference mechanism, \deeptyper, was proposed~\cite{deeptyper}. Inspired by past work on
natural language processing, \deeptyper takes a tokenized source program as input and produces
 type tokens for identifiers. Its actual architecture employs a bi-directional gated 
recurrent unit network to capture a large context around each token. They also incorporated naming information (e.g., variable names) into the type inference.
\lambdanet~\cite{lambdanet} used a graph neural network to perform type inference. It leveraged an auxiliary analysis to help with building a dataflow-graph-like structure -- the 
 type dependence graph -- as input, which establishes the relationship among program elements.  Section~\ref{sec:expr} included a detailed comparison of the accuracy and efficiency of our approach relative to \deeptyper and \lambdanet.

The application of GNNs to program analyses can be seen in other recent work as well.
One example is the gated graph neural network (GGNN)~\cite{ggnn}, which has been used to learn program representations~\cite{Allamanis17}.
The graph used in this work is based on an extension to the abstract syntax tree by adding a few types of edges to represent extra information including lexical order and variable usages.
Their approach is applied to program property prediction tasks, e.g., variable misuse detection.
\cite{Brockschmidt19} introduced a generative code modeling approach, which is also based on GGNNs, and can be applied to program repair and variable-misuse tasks. \cite{Hellendoorn20} 
used a combination of sequence layers and graph message-passing layers in their model to capture contextual information from the input program.
By leveraging more program context information using the sequence layers, their model outperformed the earlier work in \cite{Allamanis17}.
\cite{code2inv} introduced a reinforcement learning system 
to infer loop invariants in an input program.
A GNN was used to construct the structural external memory representation for a program.

Compared with past work, our approach introduces a lightweight \gnn-based type inference system that constructs a simple graph 
structure (the type flow graph) via an efficient traversal of the abstract syntax tree.
To enhance the system's accuracy and achieve high efficiency, we evaluated several GNN design choices 
that can improve the information representation and computation efficiency, including recurrent update process, attention mechanism, and a contextual layer for improving the graph node state initialization.
Based on our neural architecture design approach, our \gnn-based system shows a strong potential to be a viable candidate for lightweight analyses with superior accuracy for use in a variety of interactive programming tasks.

\section{Conclusions and Future Work}\label{sec:conc}
In this paper, we advanced past work by introducing a range of 
graph-based deep learning models that operate on a novel \textit{type 
flow graph} (TFG) representation.   Our GNN based 
models learn to propagate type information on the graph and then predict types for the graph nodes.
We studied different design choices for our GNN-based type inference system for the 100 most common types in our evaluation corpus, and show that our best GNN configuration for accuracy (\gggnnsctx{}) achieves a top-1 accuracy of 87.76\%.  This outperforms the two most closely related deep learning type inference approaches from past work -- DeepTyper with a top-1 accuracy of 84.62\%  and LambdaNet with a top-1 accuracy of 79.45\%.  Alternatively, we can state the error (100\% - accuracy) for \gggnnsctx{} is 0.80$\times$ that of DeepTyper and 0.60$\times$ that of LambdaNet.  Further, the average inference throughput of \gggnnsctx{}  is 353.8 files/second, compared to  186.7 files/second for DeepTyper and  1,050.3 files/second for LambdaNet.  If inference throughput is a higher  priority, then the recommended model to use from our approach is the next best GNN configuration from the perspective of accuracy (\gggnns{}) which achieved a top-1 accuracy of 86.89\% and an average inference throughput of 1,303.9 files/second.
Thus, our work introduces advances in graph-based deep learning that yield superior accuracy and performance to past work on  probabilistic type analysis, while also providing a range of GNN models that could be applicable in the future to other graph structures used in program analysis beyond the TFG.

\bibliography{gnn_typeinference}

\end{document}